# Study of Pakistan Election System as "Intelligent e-Election"

Muhammad Nadeem, Dr. Javaid R. Laghari, *SZABIST*

Faculty of Computer Science and Software Engineering

SZABIST, KARACHI

Abstract— The proposed election system lies in ensuring that it is transparent and impartial. Those who rule must be true representative of people's. Thus while the electoral system may vary from country to country, It has to take into account the peculiarities of every society while at the same time incorporating remedies to problems prevailing in the system.

The Electoral process expressed serious concerns regarding the independence of the Election Commission of Pakistan, the restrictions on political parties and their candidates, the misuse of state resources, some unbalanced coverage in the state media, deficiencies in the compilation of the voting register and significant problems relating to the provision of ID cards.

The holding of a general election does not in itself guarantee the restoration of democracy. The unjustified interference with electoral arrangements, as detailed above, irrespective of the alleged motivation, resulted in serious flaws being inflicted on the electoral process. Additionally, questions remain as to whether or not there will be a full transfer of power from a military to civilian administration.

The Independent study has following modules:

- Login/Subscription Module
- Candidate Subscription Module
- Vote casting Module
- Administration Module
- Intelligent decision data analysis Module

## I. INTRODUCTION

As the world watched the electoral drama unfold in Pakistan in 2003, people started wondering, "Wouldn't all our problems be solved if they just used Internet Voting?" People all over the world soon started taking a hard look at their voting equipment and procedures, and trying to figure out how to improve them. There is a strong inclination towards moving to Remote Internet Voting – at least among the politicians – in order to enhance voter convenience, increase voter confidence and voter turnout. However, as will be seen later in this independent study, there are serious technological and social aspects that make Remote Internet Voting infeasible in the visible future. Therefore, many technologists have suggested that remote poll-site electronic voting, where the voter can vote at any poll-site (not only his home county poll-site), seems to be the best step forward as it provides better voter convenience, but at the same time, does not compromise security. This paper presents a survey of the state of the art in Electronic Voting, including the various works done in Internet Voting (and the arguments against its use), as well as in electronic poll-site voting.

The state of Pakistan emerged from the partition of the Indian sub-continent in 1947. Originally created to meet the demands of Indian Muslims for their own homeland, it split in two in 1971 when the predominantly Bengali speaking eastern part of the country seceded with help from India, to become Bangladesh **[1].** With a population of 142 million, it is the seventh largest country in the world. Since independence in 1947, alternating periods of civilian and military rule and human rights abuses have undermined political and economic stability in Pakistan. During the last few decades' corruption, inefficiency and confrontations between internal institutions have tarnished civilian politics and there has been a series of military coups **[2].**

The most recent of these occurred in October 1999 when the elected government of Prime Minister Nawaz Sharif was overthrown in a bloodless coup by General Pervez Musharraf **[3]**. In accordance with this judgment, on 14th August 2001, General Musharraf announced a 'roadmap for the restoration of democracy' indicating that elections to the Provincial and National Assemblies as well the Senate would be held in October 2002. Prior to this, on 30th April 2002 President Musharraf held a referendum seeking endorsement for an extension of his rule for a further five years **[4]**.

## II. THE ELECTORAL PROCESS

In applying electoral criteria polling day itself had gone relatively smoothly and that any shortcomings were the consequence of inadequate training and administrative arrangements rather than the consequences of intended abuse. However, it had serious misgivings regarding other aspects of the electoral process.

The Electoral process expressed serious concerns regarding the independence of the Election Commission of Pakistan, the restrictions on political parties and their candidates, the misuse of state resources, some unbalanced coverage in the state media, deficiencies in the compilation of the voting register and significant problems relating to the provision of ID cards.

The holding of a general election does not in itself guarantee the restoration of democracy. The unjustified interference with electoral arrangements, as detailed above, irrespective of the alleged motivation, resulted in serious flaws being inflicted on the electoral process. Additionally, questions still remain as to whether or not there will be a full transfer of power from a military to civilian administration.

### III. EXISTING PAKISTAN ELECTION SYSTEM

The core of election system lies in ensuring that it is transparent and impartial. Those who rule must be true representatives of the people. Thus while the electoral system may vary from country to country, it has to take into account the peculiarities of every society while at the same time incorporating remedies to problems prevailing in the system.

Pakistan is a Federal Republic and is known as the Islamic Republic of Pakistan. It came into existence on 14th August, 1947. The territories of Pakistan comprise the provinces of Punjab, Sindh, North-West Frontier and Balochistan, the Islamabad Capital Territory i.e., the Federal Capital, the Federally Administered Tribal Areas (FATAs) and such states and territories as are or may be included in Pakistan whether by accession or otherwise. It has a population of more than 132,344,546 and a territory spread over 7,96,096 square kilometers [5].

The Election Commission `of Pakistan is an independent and autonomous constitutional body charged with the function of conducting transparent, free, fair and impartial elections to the National and Provincial Assemblies. The holding of elections to the office of the President and the Senate are, however, the functions of the Chief Election Commissioner. Under the existing laws the conduct of Local Government Elections is also the responsibility of the Chief Election Commissioner [5].

E-elections is designed as a service-based system, with costs amortized over time and provision for technology upgrades.

A Proclamation of Emergency issued the same day followed the takeover on 12 October 1999 and the assumption of Chief Executive's responsibilities by General Pervez Musharraf. On 14 October 1999, the Provisional Constitutional Order No. 1 of 1999 (PCO) was promulgated laying down that: Notwithstanding the abeyance of the 1973 Constitution, Pakistan would be governed "as nearly as may be, in accordance with the [1973] Constitution, subject to the Provisional Constitution and any other Orders made by the Chief Executive" [5].

*SUPREME COURT JUDGMENT MAY, 2000*

The next step taken by the military government was in the direction of its judicial legitimization. On 12 May 2000, the Supreme Court issued a Judgment acknowledging, that "the intervention of the Armed Forces through an extra-constitutional measure became inevitable" and validating the military takeover "on the basis of the doctrine of State necessity". General Musharraf's government pledged to improve accountability by rooting out corruption, to introduce institutional reforms, by devolving power to the local authorities, and hold general elections at the end of the three-year period [6].

*Alliances*

Given the large number of political parties and independent candidates running in the elections, the fragmentation of the political landscape was perhaps the most pronounced feature of Pakistan's domestic politics in the run-up to 10 October 2002.

Many parties had split into numerous factions under the same name (e.g. twelve different factions in the case of the Pakistan Muslim League - PML), headed by politicians with irreconcilable personal agendas rather than true ideological differences [7].

Apart from the fragmentation, the electoral system ("first-past-the-post") made projections extremely difficult. At the same time, one witnessed some political "marriages of convenience" that would have seemed unthinkable until recently.

### IV. ELECTORAL SYSTEM

In early 2002, the military government suggested revised electoral systems to both the Parliament and the Provincial Assemblies. After consultations with civil society organizations, political parties and other stakeholders General Musharraf finally announced the changes to the electoral system on 21 August 2002 as the Legal Framework Order, 2002 was introduced. It was also announced that elections to the National and Provincial Assemblies would take place on 10 October 2002 while elections to the Senate were re-scheduled for 12 November 2002.

The more salient changes were:

- Number of seats in the National Assembly increased;
- Number of seats in the Senate increased;
- Number of seats in the Provincial Assemblies increased;
- Set aside seats for women were introduced in both houses of the Parliament and the Provincial Assemblies;
- Separate electorate for the minority seats was abolished and a joint electorate
- introduced;[1]
- Voting age was reduced from 21 years of age to 18.

The electoral system of Pakistan reflects the federal system laid out in the 1973 Constitution. As a result, the provinces have a significant impact as how the Senate is composed, which is the norm in most federations.

---

[1] Previously non-Muslims who where registered voters constituted a separate electorate and cast their vote in direct elections for the non-Muslim seats.

All the three assemblies have a mix of directly and indirectly elected representatives. A vast majority of the members of the National Assembly and the Provincial Assemblies are directly elected (79% and 78% respectively) while the opposite relationship can be found in the Senate (92% indirectly elected by the Provincial Assemblies and 8% directly elected by the FATA electorate)[2].

*NATIONAL ASSEMBLY*

The National Assembly has three categories of seats; (1) general seats, (2) women seats and (3) non-Muslim seats. It should be stressed that only the general seats are directly elected via the First-Past-The-Post system, i.e. by simple majority in single member districts. The women seats, on the other hand, are indirectly elected using a proportional system based on the number of general seats won by the each political party from the Province concerned in the National Assembly. The non-Muslim seats, on the other hand, are indirectly elected via the same proportional electoral system as used for the women seats, except that the entire country constitutes the constituency. Both women and non-Muslims are picked from the respective closed party list. As a result, independent candidates could only run for the general seats and not for any of the reserved seats in the National Assembly.

National Assembly Composition: 2002 elections compared to 1997

| Category of Seats | 1997 Elections | 2002 Elections |
|---|---|---|
| General Seats | 207 | 272 |
| Women Seats | 0 | 60 |
| Non-Muslim Seats | 10 | 10 |
| Total No of Seats | 217 | 342 |

Sources: Legal Framework Order, 2002 and The Constitution of Pakistan, 1973.

A further novelty in this year's election is the five per cent threshold introduced to prevent smaller parties access to the reserved seats for women and non-Muslims. If a party doesn't receive at least five per cent of the general seats in the National Assembly, it will not be allotted any of these reserved seats to the National Assembly.

This made it more difficult for regional parties to obtain reserved seats, and as a result only MMA, PML(QA) and the PPPP gained access to the set aside seats. The number of seats in the National Assembly (NA) increased significantly. In 1997 elections, 217 members of the NA were elected (directly and indirectly) while this time around the figure had risen to 342. One of the main reasons for this increase was the introduction of dedicated seats for women (60), but also the general seats increased by sixty-five new seats. Also after the increase in the number of seats in the National Assembly, Punjab will still be the key province since it holds more than fifty per cent of the seats (see table below).

---

[2] It is yet to be decided how the four seats representing the Islamabad Federal Capital (IFC) should be filled in the Senate. It's the prerogative of the President to make this decision.

National Assembly seats by provinces and other administrative areas

| Geographical Area | 2002 Elections | | | Total | Percent of total Seats |
|---|---|---|---|---|---|
| | General | Women | Non-Muslim | | |
| Punjab | 148 | 35 | | 183 | 53,5% |
| Sindh | 61 | 14 | | 75 | 22% |
| Balochistan | 14 | 3 | | 17 | 5% |
| N.W.F.P | 35 | 8 | | 43 | 12,5% |
| F.A.T.A. | 12 | - | | 12 | 3,5% |
| IFC | 2 | | | 2 | 0,5% |
| Nation-wide | | | 10 | | 3% |
| Total Seats | 272 | 60 | 10 | 342 | 100% |

Source: Legal Framework Order, 2002

*SENATE*

However, as the newly elected members of the respective Provincial Assemblies cast a secret vote, there exists both room for political maneuvering and political pressure. All members of the Senate, except FATA seats, are indirectly elected. To date it has not yet been decided how the four Senate seats representing Islamabad Federal Capital shall be elected. Members to General, Women and Technocrat seats will all be elected via a proportional representation (Single Transferable Vote system). The newly elected members of the Provincial Assemblies will elect the Senators representing their respective province in the Senate based on the closed list of candidates supplied by the political parties prior to the poll.

## V. OVERVIEW OF PROPOSED SYSTEM

Voting refers to the use of computers or computerized voting equipment to cast ballots in an election. Sometimes, this term is used more specifically to refer to voting that takes place over the Internet. Electronic systems can be used to register voters, tally ballots, and record votes **[8,9]**.

The study will have following modules.

1. Login/Subscription Module
2. Candidate Subscription Module
3. Vote casting Module
4. Administration Module
5. Intelligent decision data analysis

*LOGIN/SYSTEM CONFIGURATION MODULE*

This module will provide certain overviews and brief users on rules and regulations pertaining to the system. It will facilitate the user in getting familiarized with the system. Various options will be available to the user in this section, namely, live video presentation of candidates' debates, chat-rooms for discussion with other users and forums through which participants with common interests can exchange open messages. User may provide certain details in order to receive election results via SMS or e-mail.

A registration process will be incorporated in this module to enable the users to get them registered in order to vote online.

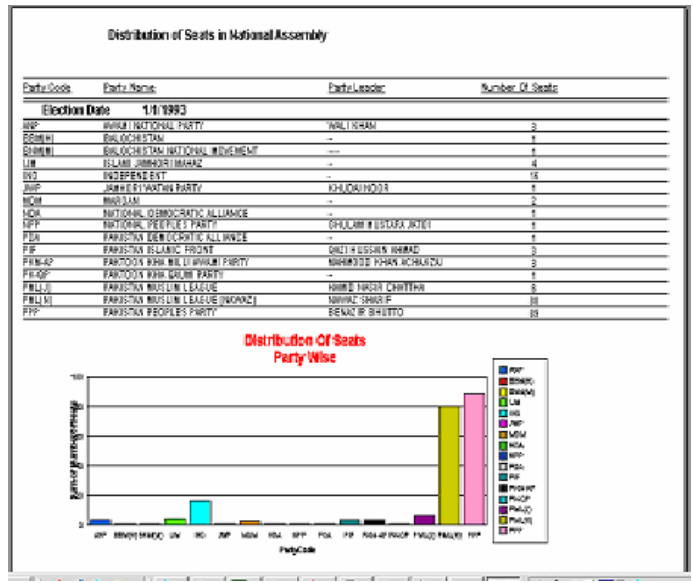

**Fig 1: Intelligent Reporting Tool (Selected National Assembly 1993 - 2002 Election Data Party Wise distribution in Last "N" Elections)**

*CANDIDATES SUBSCRIPTION MODULE*

This is a separate module that would be available online few days prior to the actual e-election date. The relevant legal provisions regarding political parties are available in the Political Parties Act, 1962 and the Political Parties Rules, 1986. Political Parties are not required to register with the Election Commission. This is in compliance with the decision of the Supreme Court of Pakistan in the case reported as PLD 1988 SC 416. It was held by the Supreme Court that the registration of Political Parties is against the Fundamental Rights enshrined in the Constitution. During the 1997 general elections, the Election Commission allocated symbols to 62 Muslim Political Parties and 15 non-Muslim Political Parties.

After permission from the Election Commission all candidates who wish to contest the elections would be required to fill the online subscription form providing the following details.

1. Academics
2. Personal Information
3. Voting symbol
4. Party name and other relevant particulars
5. Region
6. Assets etc…

The candidates will again go through a verification process which will decide as to whether they are eligible to stand for these elections or not.

*VOTE CASTING MODULE*

This module would allow Pakistani citizens all around the globe to exercise their right to vote on the day of election.

To enter into the e-elections software the user would need to login after which a screen would welcome the user where he would be required to enter his national ID card number and serial number provided to him by national identity card database. At this stage, the user ID card number and serial number would be verified by the ID card database to welcome only Pakistani citizens. For administrative purpose the administrators will be assigned unique IDs and passwords by the system developers using which they would log into the system.

Each user would log into the system through his National Identity Card number and serial numbers. These ID card numbers and serial numbers will be authenticated by the system using existing ID cards database to ensure verified users.

The user can now vote for his chosen party member. A screen would be displayed with information about all the candidates among whom the user will make a choice.

The choices would be for both the national and provincial assembly contestants.

*SECURITY*

Our proposed system would lead to a reform in the current election system by offering officials highly accurate and reliable voting platform for public elections. Our software is truly groundbreaking. Election officials won't have to sacrifice ease of use for security and privacy because it addresses the required criteria for secure, private, verifiable elections packaged in an affordable, scalable, and fully transparent voting system.

To ensure overall security of the system, SSL will be utilized. SSL works by using a public key to encrypt data that's transferred over the SSL connection. Both Netscape Navigator and Internet Explorer support SSL and many Web sites use the protocol to obtain confidential user information, such as credit card numbers. By convention, URLs that require an SSL connection start with https: instead of http: "Electronic Voting" [10], rivet addresses some issues like the "secure platform problem" and the impossibility of giving a receipt to the voter. It also provides some personal opinions on a host of issues including the striking dissimilarity between e-commerce and e-voting, the dangers of adversaries performing automated, wide-scale attacks while voting from home, the need for extreme simplicity of voting equipment, the importance of audit-trails, support for disabled voters, security problems of absentee ballots, etc.

The SSL protocol runs above TCP/IP and below higher-level protocols such as HTTP or IMAP. It uses TCP/IP on behalf of the higher-level protocols, and in the process allows an SSL-enabled server to authenticate itself to an SSL-enabled client, allows the client to authenticate itself to the server, and allows both machines to establish an encrypted connection. These capabilities address fundamental concerns about communication over the Internet and other TCP/IP networks:

SSL server authentication allows a user to confirm a server's identity. SSL-enabled client software can use standard techniques of public-key cryptography to check that a server's certificate and public ID are valid and have been issued by a certificate authority (CA) listed in the client's list of trusted CAs. This confirmation might be important if the user, for example, is sending a credit card number over the network and wants to check the receiving server's identity.

SSL client authentication allows a server to confirm a user's identity. Using the same techniques as those used for server authentication, SSL-enabled server software can check that a client's certificate and public ID are valid and have been issued by a certificate authority (CA) listed in the server's list of trusted CAs. This confirmation might be important if the server, for example, is a bank sending confidential financial information to a customer and wants to check the recipient's identity.

Discussion on requirements, threat perceptions, and socio-political issues regarding electronic voting can be found in [11, 12, 13, 14, and 15].

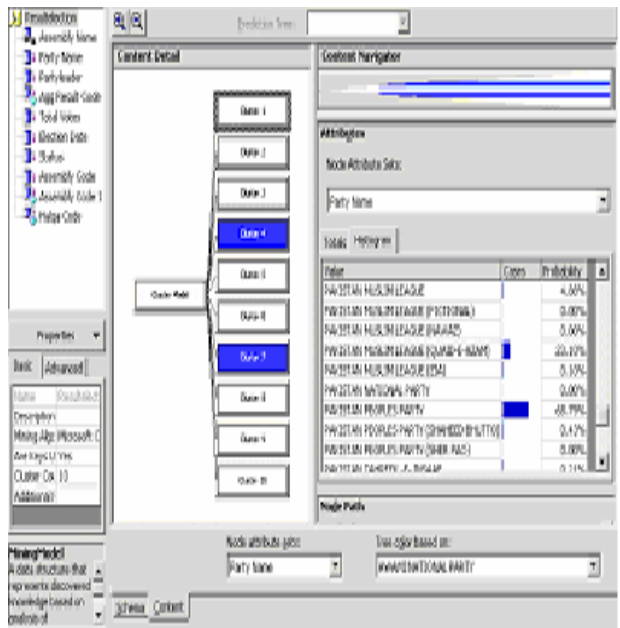
**Fig 2: Histogram cases (Party wise data)**

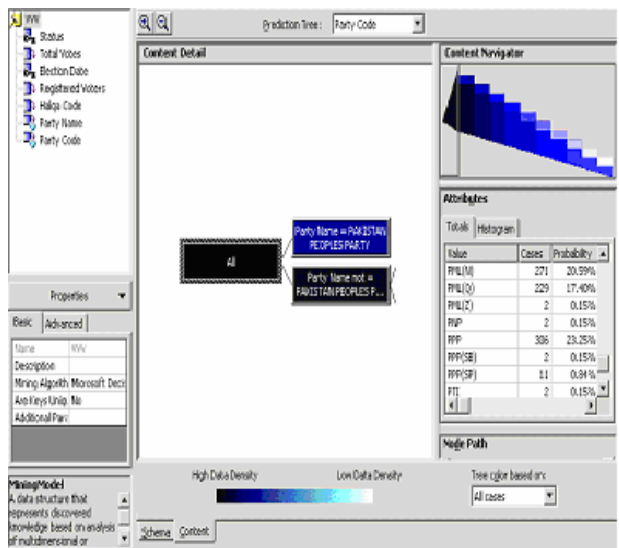
**Fig 3: Election System: Intelligent Decision Tree (All Cases)**

## VI. CONCLUSION

As the world watched the electoral drama unfold in Pakistan in 2002, people started wondering, "Wouldn't all our problems be solved if they just used Internet Voting?" People all over the world soon started taking a hard look at their voting equipment and procedures, and trying to figure out how to improve them. There is a strong inclination towards moving to Remote Internet Voting – at least among the politicians – in order to enhance voter convenience, increase voter confidence and voter turnout. However, as will be seen later in this independent study, there are serious technological and social aspects that make Remote Internet Voting infeasible in the visible future. Therefore, many technologists have suggested that remote poll-site electronic voting, where the voter can vote at any poll-site (not only his home poll-site), seems to be the best step forward as it provides better voter convenience, but at the same time, does not compromise security. This paper presents a survey of the state of the art in Electronic Voting, including the various works done in Internet Voting (and the arguments against its use), as well as in electronic poll-site voting.

## VII. ACKNOWLEDGMENT

By the grace of Allah, this independent study stands complete. However, calling it independent is not quite correct. I owe the success of this effort to my institute and my Independent study supervisor, Dr. Javaid R. Laghari, without whose guidance this accomplishment would have been impossible. Right from the very start when he accepted to being my supervisor he has been a source of inspiration and I can state that he helped tremendously in making it a far better study report than it would otherwise have been. His helpfulness was apparent in numerous ways. He contributed his infectious enthusiasm and encouragement and took time out of busy days to talk to me whenever required. The Center of Information & Research (SZABIST) was also an immense resource for providing study material that really helped me out in using the latest technology, which I have used to practically implement my study.

Above all, any remaining errors, impertinencies and imbecilities are entirely my responsibility.

This study is dedicated to all those who have given their great support.